\documentclass[aoas,MSNbibl,nameyear,dvips]{arximspdf}
\usepackage{dcolumn}
\usepackage{graphicx}
\doi{10.1214/12-AOAS624} 
\volume{7}
\issue{3}
\pubyear{2013}
\firstpage{1362}
\lastpage{1385}

\makeatletter
\newcolumntype{d}[1]{D{.}{.}{#1}}
\newcommand\postsep{\vspace{-1.3mm}}
\newcommand\M{M}
\newcommand\mort{m}
\newcommand\U{U}
\newcommand\HP{{\mathcal{H}}}

\newcommand{\no}{no}
\newcommand{\yes}{yes}

\newcommand\betaU{\alpha}
\newcommand\betaN{\beta}
\newcommand\betaA{\gamma}
\newcommand\betaY{\delta}
\newcommand\betaS{\phi}
\newcommand\betaTrend{\psi}

\makeatother

\begin{document}
\begin{frontmatter}

\title{Macroeconomic effects on mortality revealed by panel analysis
with nonlinear trends\thanksref{T1}}
\runtitle{Macroeconomic effects on mortality}
\thankstext{T1}{Supported by NIH/NICHD Grant HD057411-02.}

\begin{aug}
\author[A]{\fnms{Edward L.} \snm{Ionides}\corref{}\ead[label=e1]{ionides@umich.edu}},
\author[A]{\fnms{Zhen} \snm{Wang}\ead[label=e2]{wzhen@umich.edu}}
\and
\author[B]{\fnms{Jos\'{e} A.} \snm{Tapia Granados}\ead[label=e3]{jatapia@umich.edu}}
\runauthor{E.~L. Ionides, Z. Wang and J.~A. Tapia Granados}
\affiliation{University of Michigan}
\address[A]{E. L. Ionides\\
Z. Wang\\
Department of Statistics\\
University of Michigan\\
Ann Arbor, Michigan 48109-1107\\
USA\\
\printead{e1}\\
\phantom{E-mail:\ }\printead*{e2}} 
\address[B]{J.~A. Tapia Granados\\
Institute for Social Research\\
University of Michigan\\
Ann Arbor, Michigan 48104-1248\\
USA\\
\printead{e3}}
\end{aug}

\received{\smonth{10} \syear{2011}}
\revised{\smonth{12} \syear{2012}}

%
\begin{abstract}
Many investigations have used panel methods to study the relationships
between fluctuations in economic activity and mortality. A broad
consensus has emerged on the overall procyclical nature of mortality:
perhaps counter-intuitively, mortality typically rises above its trend
during expansions. This consensus has been tarnished by inconsistent
reports on the specific age groups and mortality causes involved. We
show that these inconsistencies result, in part, from the trend
specifications used in previous panel models. Standard econometric
panel analysis involves fitting regression models using ordinary least
squares, employing standard errors which are robust to temporal
autocorrelation. The model specifications include a fixed effect, and
possibly a linear trend, for each time series in the panel. We propose
alternative methodology based on nonlinear detrending. Applying our
methodology on data for the 50 US states from 1980 to 2006, we obtain
more precise and consistent results than previous studies. We find
procyclical mortality in all age groups. We find clear procyclical
mortality due to respiratory disease and traffic injuries.
Predominantly procyclical cardiovascular disease mortality and
countercyclical suicide are subject to substantial state-to-state
variation. Neither cancer nor homicide have significant macroeconomic
association.
\end{abstract}

%
\begin{keyword}
\kwd{Mortality}
\kwd{health economics}
\kwd{model misspecification}
\kwd{panel data}
\end{keyword}

\end{frontmatter}

\section{Introduction}\label{sec:intro}
The impact of fluctuations in economic activity on mortality has been a
long-running debate.
Early evidence for procyclical mortality (i.e., increased mortality
during economic booms) was presented by \citet{ogburn22}.
Subsequently, Harvey Brenner made determined efforts to support the
hypothesis of counter-cyclical mortality [e.g., \citet{brenner79}],
although his controversial statistical methods were eventually
discredited [\citet{gravelle81,wagstaff85}].
There is now evidence for procyclical mortality in many developed and
developing countries [reviewed by \citet{tapia11}].
Mortality is the most clear-cut measure of population health, but may
be the tip of an iceberg of procyclical morbidity.
Indeed, corresponding patterns have been found for other health-related
outcomes [Ruhm (\citeyear{ruhm03,ruhm05-jhe})], though these are complicated both
by the scope of available data and by the possibility of macroeconomic
influences on data collection.

Cyclical mortality is distinct from the debated relationship between
long-term economic development and long-term improvements in public health.
Nevertheless, the two debates are related: inasmuch as cyclical
mortality is observed for macroeconomic fluctuations at all time
scales, it plays a role in determining the long time scale variations
which are identified as trends.
Certainly, many factors other than macroeconomic considerations
contribute to population mortality levels [\citet{cutler06}].

Population level associations are distinct from the health consequences
of economic fluctuations on specific vulnerable groups, such as those
who become unemployed.
Adverse health outcomes are certainly associated with unemployment,
with evidence for causation in both directions [\citet{mcdonough01}].
The present investigation concerns aggregate effects, which may include
both beneficial and harmful consequences for different subpopulations.

A landmark in the investigation of cyclical mortality was the
application of panel methods by \citet{ruhm00}, allowing the
consideration of extensive spatiotemporal data.
\citet{ruhm00} analyzed annual statistics for 50 US states over 20
years and found predominantly procyclical mortality.
This conclusion has been largely confirmed by subsequent panel
investigations [Ruhm (\citeyear
{ruhm03,ruhm06,ruhm07}), \citet{neumayer04,tapia05,gerdtham06,buchmueller07,miller09}, Gonzalez and Quast (\citeyear{gonzalez10,gonzalez11})].
The spatial units in these studies vary (states, countries, regions,
French departments), but we will consistently refer to them as states.
These panel studies were typically carried out in the spirit of
difference-in-difference analysis [\citet{bertrand04}].
In this paradigm, temporal variations in mortality are controlled by
taking a difference between state mortality and national mortality,
interpreted in regression models as a fixed year effect;
spatial variations in mortality are controlled by including
state-specific mortality effects.
The resulting relationships identified between macroeconomic variables
and mortality are therefore resistant to bias from either strictly
spatial or strictly temporal additive omitted variables.
By removing national mortality effects, difference-in-difference panel
analysis is complementary to time series analysis [\citet{ruhm05}],
though the two approaches have led to broadly consistent results [\citet
{tapia05-ije}].
Individual-level data have also revealed predominantly procyclical
effects [\citet{edwards08}].
Sample size issues limit the scope of individual-level analyses;
macroeconomic impact on mortality is an effect of small size (for any
given individual), which nevertheless has a considerable overall effect
on entire populations.

A critical question for the proper understanding of procyclical
aggregate mortality is to what extent different age groups and
mortality causes are involved in the procyclical phenomenon.
Problematically, different analyses have previously led to different answers.
We argue that these inconsistencies can be explained by the use of
misspecified state-specific trend models.
Previous analyses have typically employed linear or constant
state-specific trends and have performed statistical regression
techniques which are inefficient or biased for the data under consideration.
The methodological limitations of these analyses have had severe
consequences for investigating age and cause-specific mortality,
without being large enough to interfere with the results for total mortality.
To support our argument, we will show how removal of nonlinear trends
allows appropriate statistical analysis using standard regression methods.

In this investigation, we study data from the US states in the years
$1980\mbox{--}2006$.
Thus, our data updates the 1972--1991 analysis of \citet{ruhm00} and
overlaps the 1978--2004 analysis of \citet{miller09}.
Whereas \citet{miller09} extended \citet{ruhm00} by breaking down the
data more extensively by age and mortality cause, here we focus instead
on the specification of the model and its consequences for the
conclusions reached.
We find that some estimates of interest are fragile to changes in the
specification.
Results which are sensitive to the model specification should be
treated with additional caution and also raise the question of which
specification is most appropriate.
To resolve existing ambiguities, and to make further progress, there is
a need for objective evaluation of the strengths and weaknesses of
alternative analyses.
Assessing the model specification via analysis of the regression
residuals can provide such a tool.
The constant or linear state-specific trend specifications used in
previous work, including \citet{ruhm00} and \citet{miller09}, entail
substantial violations of the standard assumptions that justify the use
of ordinary least squares (OLS).
Combining OLS point estimates with state-clustered standard errors is a
standard econometric technique in this situation [\citet
{bertrand04,petersen09}], however, this only partially alleviates the
adverse consequences of the model violations.
Our methodological remedy is to apply nonlinear detrending methods in
this spatiotemporal setting.
We show that our method has many of the advantages of feasible
generalized least squares (FGLS) while avoiding some of the
difficulties inherent in using data to estimate a large covariance
matrix [\citet{hausman08}].

Our results confirm the finding of \citet{ruhm00} that general
mortality fluctuates procyclically and this procylical phenomenon is
stronger in young adults (ages 20--44), though it is present also in
mid-age adults (45--64) and individuals at retirement ages (65$+$).
The conclusion of \citet{miller09} that mid-age adults are not subject
to procyclical mortality may be a consequence of model misspecification.
Since \citet{miller09} and \citet{ruhm00} used similar model
formulations, it is fortuitous that many of the results of \citet
{ruhm00} happen to agree with the conclusions from a more statistically
principled analysis of recent data.
We find that the procyclical oscillation of general mortality is mainly
mediated by increases in respiratory disease mortality, cardiovascular
disease mortality and traffic mortality, all of which oscillate procyclically.
Suicide differs by being countercyclical; we find cancer and homicide
to be acyclical.

The remainder of this paper is organized as follows. Section~\ref
{sec:data} describes the data.
Section~\ref{sec:models} introduces the panel models under consideration.
Section~\ref{sec:methods} discusses the methodological issues involved
in fitting these models.
Section~\ref{sec:results} carries out a data analysis, focusing on
issues of methodological relevance.
Section~\ref{sec:diagnostics} investigates goodness of fit for the
models under consideration.
Section~\ref{sec:discussion} discusses these results in the context of
the current understanding of cyclical mortality.

\section{Data}\label{sec:data}

We analyzed annual data from the $50$ US states over the years
$1980\mbox{--}2006$.
Crude, age-specific, sex-specific and cause-specific mortality rates
were computed using data publicly available from the US Centers for
Disease Control and Prevention (\href{http://wonder.cdc.gov}{wonder.cdc.gov}).
Data on annual unemployment rates were obtained from the US Bureau of
Labor Statistics (\href{http://www.bls.gov}{www.bls.gov}).
Age-specific mortality rates were calculated for three age groups:
$20\mbox{--}44$, $45\mbox{--}64$ and 65$+$. We analyzed cause-specific
mortality rates for eight major causes of death, defined via (ICD9;
ICD10) codes as follows: cardiovascular disease (390--459; I00--I99),
ischemic heart disease (410--414; I20--I25), cancer (140--165,
170--175, 179--203; C00--C97), respiratory disease (460--519;
J00--J98), other infectious disease (001--139; A00--B99), traffic
injuries (E810--E819; V01--V79), suicide (E950--E959; X60--X84),
homicide (E960--E969; X85--Y09).

Inspection of the plotted series of mortality rates for the 50 states
revealed a jump in the series for ischemic heart disease and cancer
mortality between the years 1998 and 1999 (results not shown) which
corresponds to the transition in disease coding from the 9th to the
10th edition of the International Classification of Diseases (i.e.,
from ICD9 to ICD10).
The annual change in ischemic heart disease mortality took its largest
value at this time for 48 states.
For cancer, the largest annual change occurred at this time for 20
states, with the times of the biggest jump being scattered for the
other states.
To correct for the potential error introduced by a change in mortality
codes for these two categories, we replaced the log mortality increment
for 1998--1999 by the average value of the remaining increments (a~simple way to remove the effect of the jump while keeping the temporal
structure of the time series intact).
This data correction made no qualitative difference to our conclusions.

\section{Models}\label{sec:models}

We consider panel model specifications extending the choices of \citet
{ruhm00}. Our general model is
%
%
\begin{equation}
\M_{it}= \betaU\U_{it}+ \betaN N_t + \betaA
A_{it} + \betaY_t + \betaS_i +
\betaTrend_{i}t + \varepsilon_{it}, \label{eq:model}
\end{equation}
where $\M_{it}$ is a measure of mortality for state $i$ in year $t$;
$\U
_{it}$ is a measure of state-level unemployment;
$N_{t}$ is a measure of national unemployment; $A_{it}$ is a column
vector representing population age-structure,\setcounter{footnote}{1}\footnote{Age-adjusted
state mortality rates are available.
However, \citet{rosenbaum84} have pointed out the potential biases
introduced by using age-adjusted rates. Following these authors, we
prefer to regress crude rates on a set of covariates including
age-structure variables.}
with $\betaA$ being a row vector of corresponding size;
$\betaY_t$ are year-specific state-invariant effects;
$\betaS_i$ are state-specific time-invariant effects;
the term $\betaTrend_{i}t$ corresponds to state-specific linear trends;
$\varepsilon_{it}$ is an error term.
The mortality rate measure, $\M_{it}$, may correspond to total
mortality, age-specific mortality or cause-specific mortality.
When $\M_{it}$ is an age-specific mortality measure, we do not include
the term $\betaA A_{it}$.

To define our mortality and unemployment measures, we introduce
notation for the raw data.
The mortality rate data are denoted by $\mort_{it}$, state-specific
unemployment rate by $u_{it}$ and national unemployment rate by $n_t$.
A~vector with population proportions of children under 5 and of persons
over 65 is written as $a_{it}$.
We consider four types of model, corresponding to four different ways
to work with state-specific levels and trends:
\begin{longlist}[(HP$_\lambda$)]
\item[(B)]
The \textit{Basic} model is the foundation for the analysis of \citet{ruhm00}.
It has dependent variable $\M_{it}=\log m_{it}$ and fits a constant
level effect for each state (i.e., it has a constraint $\betaTrend_i=0$).
The remaining variables are untransformed ($U_{it}=u_{it}$, $N_t=n_t$,
$A_{it}=a_{it}$).

\item[(L)]
The \textit{Linear} model includes linear state-specific trends.
The linear model differs from the basic model only by the inclusion of
the term $\betaTrend_i t$.
\item[(D)]

The \textit{Differenced} model includes all time-dependent variables in
first temporal differences.
Specifically, $\M_{it}=\Delta\log m_{it}=m_{it+1}-m_{it}$, $\U
_{it}=\Delta u_{it}$, $N_{t}=\Delta n_{t}$, and $A_{it}=\Delta a_{it}$.

\item[(HP$_\lambda$)]
The \textit{Hodrick--Prescott} model includes the time-dependent variables
after subtracting trends computed via a Hodrick--Prescott filter with
smoothing parameter $\lambda$.
In this case, we write $\M_{it}=\HP_\lambda(\log m_{it})$, $\U
_{it}=\HP
_\lambda(u_{it})$, $N_{t}=\HP_\lambda(n_{t})$, and $A_{it}=\HP
_\lambda(a_{it})$.
Here, $\HP_\lambda(x_t)$ denotes the residual component of the time
series $x_t$ after removing a trend computed by the method of \citet
{hodrick97}.
As discussed in Section~\ref{sec:methods}, and at greater length by
\citet{ionides12-aoas-sup}, $\lambda$ can be chosen to approximately
prewhiten the mortality measure rather than aiming specifically to
isolate business cycle fluctuations.
The choice $\lambda=100$ satisfies this requirement [\citet
{ionides12-aoas-sup}, Figure~S-2].
\end{longlist}

The model types are summarized in Table~\ref{tab:models}(a).
All regression models were weighted by the square root of the state
population to account for heteroskedasticity;
this has become a standard formulation [\citet
{ruhm00,tapia05,miller09,gonzalez11}].
State-specific fixed effects and linear trends are removed by the
Hodrick--Prescott filter and so are not included in models of type
HP$_\lambda$.
The linear trends in models of type L correspond to fixed effects after
temporal differencing; we therefore include state-specific fixed
effects in models of type~D.

%
%
\begin{table}
\caption{Models under consideration, written as special cases of
equation (\protect\ref{eq:model}). \textup{(a)} The mortality variable and
time-dependent explanatory variables for the different model types.
\textup{(b)} Model subtypes, including differing subsets of the
explanatory variables}\label{tab:models}
\begin{tabular*}{\textwidth}{@{\extracolsep{\fill}}lcccccc@{}}
\multicolumn{7}{c}{(a)}\\
\hline
&\textbf{Mortality}&\textbf{State}&\textbf{National}&\textbf
{Age}&\textbf{Fixed}&\multicolumn{1}{c@{}}{\textbf{Linear}}\\
\textbf{Model} &\textbf{measure}&\textbf{economy}&\textbf
{economy}&\textbf{structure}&\textbf{effect}&\multicolumn
{1}{c@{}}{\textbf{trend}}\\
\textbf{type}&
\multicolumn{1}{c}{$\bolds{(\M_{it})}$}&
\multicolumn{1}{c}{$\bolds{(U_{it})}$}&
\multicolumn{1}{c}{$\bolds{(N_t)}$}&
\multicolumn{1}{c}{$\bolds{(A_{it})}$}&
\multicolumn{1}{c}{$\bolds{(\betaS_i)}$}&
\multicolumn{1}{c@{}}{$\bolds{(\betaTrend_{i}t)}$}
\\
\hline
B (Basic)& $\log\mort_{it}$ & $u_{it}$ & $n_t$ & $a_{it}$& \yes&\no
\\
L (Linear)& $\log\mort_{it}$ & $u_{it}$ & $n_t$ & $a_{it}$&\yes&\yes
\\
D (Difference)& $\Delta\log\mort_{it}$ & $\Delta u_{it}$ &$\Delta
n_t$ & $\Delta a_{it}$ &\no&\no\\
HP$_\lambda$ (HP-detrended)& $\HP_\lambda(\log\mort_{it})$ & $\HP
_\lambda(u_{it})$ & $\HP_\lambda(n_{t})$ & $\HP_\lambda(a_{it})$
&\no
&\no\\
\hline
\end{tabular*}
\begin{tabular*}{250pt}{@{\extracolsep{\fill}}lccc@{}}
\multicolumn{4}{c}{(b)}\\
\hline
\textbf{Model}& \textbf{State}&\textbf{National}&\textbf{Year}\\
\textbf{subtype} &\textbf{economy}&\textbf{economy}&\textbf{effects}\\
\hline
1 & \yes& \no& \yes\\
2 & \yes& \no& \no\\
3 & \no& \yes& \no\\
4 & \yes& \yes& \no\\
\hline
\end{tabular*}
\end{table}

We consider four subtypes of each model type, corresponding to the
inclusion of differing sets of covariates.
The national economy covariate, $N_t$, can be expressed as a linear
combination of the year effects, $\{\betaY_t\}$, and so we never
include both in the model simultaneously.
Model B1 has $\betaN=0$, excluding an explicit role for the national economy;
model B2 excludes both national unemployment and year effects ($\betaN
=\betaY_t=0$);
model B3 excludes state unemployment and year effects ($\betaU=\betaY_t=0$);
model B4 excludes year effects ($\betaY_t=0$).
These model subtypes were considered by \citet{ruhm00}, with the goal
of disentangling the effects of state-level unemployment and
national-level unemployment on mortality.
Subtypes of the other model types are defined in an identical way, as
summarized in Table~\ref{tab:models}(b).


\section{Methodology}\label{sec:methods}

In a panel study such as ours, many variables are measured at many
geographical locations across many time points.
This wealth of data leads to challenges in building graphical representations.
Nevertheless, plotting the data or regression residuals in various ways
can play an important role in model development and diagnostics.
We demonstrate this in Sections~\ref{sec:results} and~\ref{sec:diagnostics}.
By contrast, previous panel studies relating mortality to
macroeconomics have not shown any graphical representations of the data
below national levels of aggregation.

A classical approach to regression analysis is to present estimates and
standard errors based on OLS methodology, after checking that thorough
investigation of the residuals does not reveal any major violations of
the corresponding model assumptions.
When serious violations are discovered one seeks to remove them by
respecifying the model, for example, by using transformations of
variables or appropriately weighting the error terms.
An alternative approach to inference is to employ nonparametric error
models which operate under weaker assumptions, as discussed in the
context of panel analysis by \citet{bertrand04} and \citet{petersen09}.
A hidden cost of nonparametric error models is that the finite-sample
properties can be undesirable despite demonstrably good asymptotic
properties [\citet{kauermann01}].
In numerical experiments, a sample size of 50 independent time series
has sometimes been found sufficient to validate the asymptotic
justification of robust standard errors for panel models [\citet
{bertrand04,petersen09}].
However, the numerical validation is dependent on the data and models
under consideration and so should be rechecked on a case-by-case basis.
If a relatively simple respecification justifies standard OLS
techniques, the additional complexities of employing and validating
nonparametric error models can be avoided.

In the context of time series analysis, regression with autocorrelated
errors can be handled by a procedure called prewhitening [\citet{shumway06}].
One looks for a transformation which provides approximately
uncorrelated residuals when the transformed dependent time series is
regressed on the transformed independent series.
If the transformation has a linearity property, then the resulting OLS
estimates of the regression coefficients remain unbiased.
The linearity property of transformations is distinct from the use of
the word linear to describe the term $\beta_i t$ in equation~(\ref{eq:model}).
Transformations having this linearity property include temporal
differencing, detrending by computing the residuals from fitting an
auto-regressive moving-average model, and detrending using the
Hodrick--Prescott filter.
If application of the Hodrick--Prescott filter with a particular choice
of smoothing parameter leads to effective prewhitening, this gives a
data-driven justification of the resulting analysis.
Thus, the extensive literature on the value of the smoothing parameter
appropriate to study business cycle fluctuations in annual data [\citet
{ravn02}] is not directly relevant to our methodology.
Additional material on the interpretation and consequences of the
choice of smoothing parameter is given in the supplement [\citet
{ionides12-aoas-sup}].

Much of the development of econometric panel analysis (both in theory
and practice) has focused on the standard errors.
OLS standard errors can considerably underestimate the actual
variability of the parameter estimates, leading to great potential for
the ``discovery'' of spurious effects [\citet{bertrand04}].
A variety of methods, including clustered error estimates and bootstrap
methodology, have been proposed to amend this problem [\citet{petersen09}].
Even once the standard errors are appropriately corrected, there is a
remaining difficulty that OLS point estimates can also be unreliable in
these situations.
Feasible generalized least squares (FGLS) aims to improve OLS by using
an estimated covariance matrix for the error terms [\citet{hansen07}].
However, the use of FGLS in panel analysis is rare, amounting to just
3\% of the panel analyses surveyed by \citet{petersen09} and 1\% of
those surveyed by \citet{bertrand04}.
Applying FGLS using simple models for the covariance structure can be
ineffective [\citet{bertrand04}].
Difficulties arise in complex, flexible models of the covariance
structure due to the potentially large number of parameters to be
estimated [\citet{hansen07,hausman08}].
Our method of applying a detrending linear transform to both sides of
the regression equation~(\ref{eq:model}) is formally similar to the
application of FGLS, with detrending playing the role of covariance estimation.
From this perspective, nonlinear detrending is a variant of FGLS which
is readily interpretable and has favorable numerical properties.

\citet{granger74} encouraged analyzing temporal differences as a
practical resolution to the difficulties of jointly estimating
regression coefficients and autocovariance structures in the presence
of substantial long-range autocorrelation.
However, a relationship between differences does not readily imply a
relationship between trends: in practice, fluctuations around a trend
can have entirely different relationships to those of the trends
themselves [\citet{hodrick97}].
Temporal differencing was the only linear data transformation explored
by \citet{bertrand04}.
This transformation performed excellently in their numerical experiment
[Table~IIA, line 8 of \citet{bertrand04}].
However, the authors commented that differencing was seldom used in
their survey of current practice and gave the method no further consideration.
A concern with differencing is that it can result in substantial
negative autocorrelation of residuals (and it does so for our data).
In this case, differencing is not ideal as a prewhitening filter; it
over-enthusiastically removes the positive autocorrelation.
The typical consequences of the negative autocorrelation are
inefficient OLS effect estimates and conservative standard errors.

If trends are considered as fixed effects, rather than zero mean random
effects, then OLS and FGLS models which fail to account for these
trends incur bias.
Panel model implementations of FGLS typically assume that the error
terms are independent between states, so that only temporal
autocorrelations are substantial.
Nonlinear trends which show similarities between states are not
appropriately modeled under this assumption.
By contrast, inasmuch as these phenomena are effectively removed by a
detrending operation, the corresponding prewhitened regression is
protected from bias.
The statistical evidence in the data (Sections~\ref{sec:results}
and~\ref{sec:diagnostics}) suggests that there are unmodeled sources of
spatiotemporal dependence which can largely be removed by employing
national year effects in combination with nonlinear detrending.

Interpreting the results of observational studies requires care because
of the possible consequences of omitted variables.
Another hazard is the possibility that an association between two
variables which is interpreted as causal in one direction in fact has a
causal mechanism in the opposite direction.
In the context of cyclical mortality, two uncontroversial assertions
can assist the causal interpretation of observed associations:
\begin{longlist}[(A2)]
\item[(A1)]
It has been generally accepted that mortality fluctuations could not
plausibly be a substantial cause of recent US booms and busts.

\item[(A2)]
There is a lack of plausible noneconomic phenomena which could
simultaneously have substantial effects on civilian mortality and
macroeconomic fluctuations in recent years in the US. Perhaps the best
candidates are wars, natural disasters, climate variation, or epidemic
diseases; none of these have been previously considered as plausible
omitted variables to explain procyclical mortality.
\end{longlist}
An alternative to (A2) is to employ a broad definition of macroeconomic
phenomena, including macroeconomic effects of variables external to the
economy as well as interacting variables within the economy, by
assuming the following:
\begin{longlist}[(A3)]
\item[(A3)]
Any phenomenon with macroeconomic consequences is itself a
macroeconomic phenomenon.
\end{longlist}
If there is adequate statistical evidence for an association, then
either (A1), (A2) or (A1), (A3) implies that the association can be
interpreted as a causal effect of macroeconomic fluctuations on mortality.
This follows directly from a basic principle of inductive reasoning,
that an association between two variables must be explained by either a
direct causal effect or by each variable responding to some third
variable [\citet{mill53}].
From (A2) or (A3) one can deduce that any such third variable is itself
a macroeconomic variable.
This argument does not allow us to infer a specific causal mechanism.
In particular, we cannot infer that unemployment operates causally to
produce an observed association; its role in our analysis is as a proxy
for the multitude of economic variables (measurable and nonmeasurable)
which fluctuate synchronously.

%
\begin{figure}[b]

\includegraphics{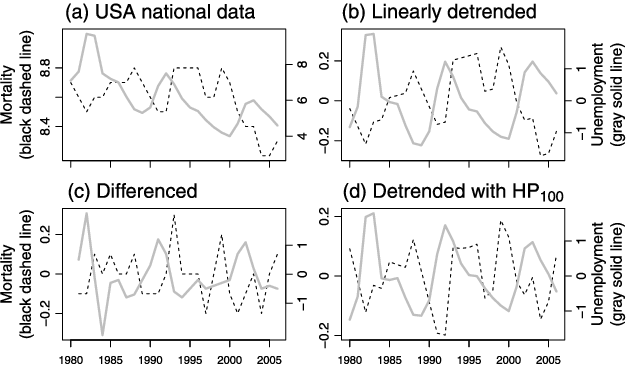}

\caption{National mortality and unemployment. \textup{(a)} Mortality per 1000
per year, shown as a dashed
line corresponding to the left axis scale;
unemployment rate, shown as a solid line corresponding to the right
axis scale. \textup{(b)}, \textup{(c)}, \textup{(d)} The
data in \textup{(a)} detrended using a linear trend, first difference and
Hodrick--Prescott filter ($\lambda=100$), respectively.}\label{fig:national}
\end{figure}

%
\begin{figure}

\includegraphics{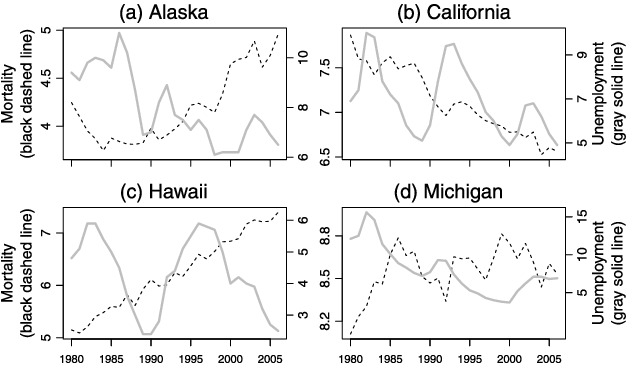}

\caption{Mortality and unemployment for four states.
Mortality per 1000 per year is shown as a dashed line corresponding to
the left axis scale.
The unemployment rate is shown as a solid gray line corresponding to
the right axis scale.}\label{fig:state}
\end{figure}

\section{Results}\label{sec:results}

Figure~\ref{fig:national} displays national annual series of total
mortality rates and the unemployment rate.
The national death rate declined dramatically during the recessions of
the early 1980s, and then increased throughout much of the expansion of
the mid-1980s.
In general, the evolution of mortality tends to mirror the evolution of
the unemployment rate, suggesting an inverse relation between
unemployment and mortality.
The long-run behavior of the crude mortality rate (unadjusted for age,
as shown in Figure~\ref{fig:national}) is affected by changes in the
age-structure of the population; it is much less likely, however, that
changes in the age-structure cause short term oscillations of the
mortality rate.
When attempting to interpret the data in Figure~\ref{fig:national}, the
strength of the statistical evidence for the association is more
critical than the issues of causal direction and omitted variable bias.
Assumptions~(A1)--(A3) can justify interpreting statistically significant
associations as macroeconomic effects on mortality, without being able
to pin down specific mechanisms.
Securing the statistical evidence in sub-categories, broken down by
demographic group and cause of mortality, then gives a foundation for
the discussion of causal mechanisms consistent with the resulting
pattern of associations.
Unfortunately, the association at the 27 annual time points in
Figure~\ref{fig:national} does not give statistically conclusive evidence.
Disaggregating mortality and unemployment rates from the national level
to the state level has potential to reinforce the evidence, as long as
the states show sufficient variation from the national pattern.
Figure~\ref{fig:state} plots mortality rates and unemployment rates for
four states, revealing quite different patterns in different states.
Some of these time series, such as mortality in Alaska, are clearly not
well modeled by variation around a linear trend.

%
\begin{table}
\tabcolsep=0pt
\caption{Fixed-effects panel regressions with state mortality modeled
as a function of economic conditions for the 50 US states}\label{tab:ruhm1}
\begin{tabular*}{\textwidth}{@{\extracolsep{\fill}}lcccccccc@{}}
%
\hline
& \multicolumn{4}{c}{\textbf{Basic model}} & \multicolumn
{4}{c@{}}{\textbf{Linear
state-specific trends}} \\[-6pt]
& \multicolumn{4}{c}{\hrulefill} & \multicolumn{4}{c@{}}{\hrulefill} \\
& \textbf{B1} & \textbf{B2} & \textbf{B3} & \textbf{B4} & \textbf{L1} &
\textbf{L2} & \textbf{L3} & \multicolumn{1}{c@{}}{\textbf{L4}} \\
\hline
State & $-0.52^{***}$ & 0.12$^\dagger$ & & $-0.67^{***}$ &
$-0.31^{***}$ & $-0.41^{***}$ & & $-0.30^{***}$ \\
unemployment & (0.07) & (0.07) & & (0.09) & (0.05) & (0.04) & & (0.06)
\\
National & & & 0.78$^{***}$ & 1.36$^{***}$ &
& & $-0.46^{***}$ & $-0.20^{**}$ \\
unemployment & & & (0.08) & (0.11) & & & (0.05) & (0.07) \\
Year effects & Yes & No & No & No & Yes & No & No & No \\
AIC & $-5818.3$ & $-5027.5$ & $-5114.7$ & $-5165.$0 & $-7569.6$ &
$-6938.5$ &
$-6917.7$ & $-6945.0$ \\
\hline
\end{tabular*}
\tabnotetext[]{}{The model specifications are as described in equation
(\ref{eq:model})
and Table~\ref{tab:models}, and were estimated using least squares,
with states weighted by the square root of the state population.
The state unemployment effect is reported as the estimate of $100\betaU
$, the percentage increase in mortality due to a unit increase in unemployment.
Similarly, the national unemployment effect is given as the estimate of
$100\betaN$.
Corresponding OLS standard errors [as used by \citet{ruhm00}] are in
parentheses.
$^{\ast\ast\ast}P<0.001$, $^{\ast\ast}P<0.01$, $^{\ast
}P<0.05$, $^{\dagger}P<0.1$.}
\end{table}

Table~\ref{tab:ruhm1} summarizes our results in models that have been
repeatedly used, following \citet{ruhm00}, to estimate the effect of
macroeconomic fluctuations on mortality.
The models with linear trends (L1--L4) give similar results to the
corresponding results for 1972--1991 [\citet{ruhm00}, Table I].
In addition, inspection of the Akaike information criterion (AIC)
values in Table~\ref{tab:ruhm1} shows that L1--L4 provide a
considerably superior statistical explanation of the data over B1--B4.
AIC is only one of many possible measures for quantitative model
comparison [\citet{burnham02}].
However, the differences between the AIC values in Table~\ref
{tab:ruhm1} are entirely unambiguous.
Differences of order 1 unit of AIC are considered small, and so
alternative methodologies might be expected to disagree; differences of
order 100 or 1000 units of AIC are compelling evidence.
The comparisons provided by these AIC values are therefore, presumably,
insensitive to the measure of model comparison used.
Differences in AIC are useful for detecting issues of model
misspecification, but they cannot,
by themselves, explain how and why this misspecification manifests itself.

\citet{ruhm00} found that B1--B4 provided qualitatively similar results
to L1--L4 and therefore proceeded to use the simpler basic
specification for subsequent analysis.
For our analysis, B1 is qualitatively consistent with L1--L4 and,
indeed, the effect estimate for this model ($-0.52$) happens to be
identical to the estimate of \citet{ruhm00}.
Problematically, B2--B4 suggest conclusions which are inconsistent both
with \citet{ruhm00} and with the other specifications in Table~\ref
{tab:ruhm1}.
Since B2--B4 also provide poor fits to the data (as judged by AIC and
diagnostic plots), this could be explained by model misspecification bias.
Model subtypes 2--4 aim to investigate the contextual role of
unemployment, addressing whether national macroeconomic conditions
continue to play a role given state-level variables.
However, our objective here is not to interpret the results from
fitting B2--B4 or L2--L4, but to observe how standard methodology can
lead to apparent contradictions.
The AIC values in Table~\ref{tab:ruhm1} suggest that year effects play
a statistically important role.
We therefore focus henceforth on models of subtype~1.

\begin{table}
\tabcolsep=4pt
\caption{Percentage increase in mortality associated with a unit
increase in the state unemployment rate, using different model
specifications}\label{tab:methods}
%
%
\begin{tabular*}{\textwidth}{@{\extracolsep{\fill}}llllll@{}}
\hline
\multicolumn{1}{@{}l}{\textbf{Model}} &
\multicolumn{1}{c}{\textbf{B1}} &
\multicolumn{1}{c}{\textbf{L1}} &
\multicolumn{1}{c}{\textbf{D1}} &
\multicolumn{1}{c}{\textbf{HP1}$\bolds{_{6.25}}$} &
\multicolumn{1}{c@{}}{\textbf
{HP1}$\bolds{_{100}}$} \\
\hline
Total & $-0.52${\scriptsize$
\begin{array}{l} \postsep{***}\\ \postsep{
\includegraphics{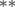}
}\\
\postsep{
\includegraphics{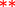}
}
\end{array}
$} & $-0.31${\scriptsize$
\begin{array}{l} \postsep{***}\\ \postsep{
\includegraphics{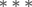}
}\\
\postsep{
\includegraphics{624i02.eps}
}
\end{array}
$} & $-0.16${\scriptsize$
\begin{array}{l} \postsep{*}\\ \postsep{
\includegraphics{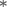}
}\\ {}
\end{array}
$} & $-0.24${\scriptsize$
\begin{array}{l} \postsep{***}\\ \postsep{
\includegraphics{624i07.eps}
}\\
\postsep{
\includegraphics{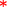}
}
\end{array}
$} & $-0.33${\scriptsize$
\begin{array}{l} \postsep{***}\\ \postsep{
\includegraphics{624i07.eps}
}\\
\postsep{
\includegraphics{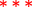}
}
\end{array}
$} \\
20--44 & $-1.15${\scriptsize$
\begin{array}{l} \postsep{***}\\ \postsep{
\includegraphics{624i06.eps}
}\\
\postsep{
\includegraphics{624i01.eps}
}
\end{array}
$} & \phantom{$-$}0.14{\scriptsize$
\begin{array}{l} \postsep{}\\ \postsep{}\\ {}
\end{array}
$} & $-0.54${\scriptsize$
\begin{array}{l} \postsep{*}\\ \postsep{
\includegraphics{624i05.eps}
}\\
\postsep{
\includegraphics{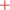}
}
\end{array}
$} & $-0.73${\scriptsize$
\begin{array}{l} \postsep{***}\\ \postsep{
\includegraphics{624i06.eps}
}\\
\postsep{
\includegraphics{624i01.eps}
}
\end{array}
$} & $-0.47${\scriptsize$
\begin{array}{l} \postsep{**}\\ \postsep{
\includegraphics{624i05.eps}
}\\ {}
\end{array}
$} \\
45--64 & $-0.72${\scriptsize$
\begin{array}{l} \postsep{***}\\ \postsep{
\includegraphics{624i06.eps}
}\\
\postsep{
\includegraphics{624i01.eps}
}
\end{array}
$} & $-0.01${\scriptsize$
\begin{array}{l} \postsep{}\\ \postsep{}\\ {}
\end{array}
$} & $-0.13${\scriptsize$
\begin{array}{l} \postsep{}\\ \postsep{}\\ {}
\end{array}
$} & $-0.14${\scriptsize$
\begin{array}{l} \postsep{}\\ \postsep{}\\ {}
\end{array}
$} & $-0.22${\scriptsize$
\begin{array}{l} \postsep{**}\\ \postsep{
\includegraphics{624i05.eps}
}\\
\postsep{
\includegraphics{624i01.eps}
}
\end{array}
$} \\
65$+$ & $-0.43${\scriptsize$
\begin{array}{l} \postsep{***}\\ \postsep{
\includegraphics{624i06.eps}
}\\
\postsep{
\includegraphics{624i01.eps}
}
\end{array}
$} & $-0.16${\scriptsize$
\begin{array}{l} \postsep{***}\\ \postsep{
\includegraphics{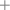}
}\\ {}
\end{array}
$} & $-0.03${\scriptsize$
\begin{array}{l} \postsep{}\\ \postsep{}\\ {}
\end{array}
$} & $-0.16${\scriptsize$
\begin{array}{l} \postsep{*}\\ \postsep{
\includegraphics{624i08.eps}
}\\ {}
\end{array}
$} & $-0.25${\scriptsize$
\begin{array}{l} \postsep{***}\\ \postsep{
\includegraphics{624i07.eps}
}\\
\postsep{
\includegraphics{624i01.eps}
}
\end{array}
$} \\
Cardiovascular disease & $-0.38${\scriptsize$
\begin{array}{l} \postsep{***}\\ \postsep{}\\ {}
\end{array}
$} & $-0.20${\scriptsize$
\begin{array}{l} \postsep{**}\\ \postsep{\hspace*{-0.5pt}
\includegraphics{624i08.eps}
}\\
\postsep{
\includegraphics{624i04.eps}
}
\end{array}
$} & $-0.06${\scriptsize$
\begin{array}{l} \postsep{}\\ \postsep{}\\ {}
\end{array}
$} & $-0.14${\scriptsize$
\begin{array}{l} \postsep{}\\ \postsep{}\\ {}
\end{array}
$} & $-0.24${\scriptsize$
\begin{array}{l} \postsep{***}\\ \postsep{
\includegraphics{624i06.eps}
}\\
\postsep{
\includegraphics{624i01.eps}
}
\end{array}
$} \\
Ischemic heart disease & $-0.33${\scriptsize$
\begin{array}{l} \postsep{\tiny{+}}\\ \postsep{}\\ {}
\end{array}
$} & $-0.35${\scriptsize$
\begin{array}{l} \postsep{**}\\ \postsep{
\includegraphics{624i08.eps}
}\\ {}
\end{array}
$} & $-0.14${\scriptsize$
\begin{array}{l} \postsep{}\\ \postsep{}\\ {}
\end{array}
$} & $-0.28${\scriptsize$
\begin{array}{l} \postsep{\tiny{+}}\\ \postsep{
\includegraphics{624i08.eps}
}\\ {}
\end{array}
$} & $-0.58${\scriptsize$
\begin{array}{l} \postsep{***}\\ \postsep{
\includegraphics{624i06.eps}
}\\
\postsep{
\includegraphics{624i01.eps}
}
\end{array}
$} \\
Cancer & $-0.20${\scriptsize$
\begin{array}{l} \postsep{*}\\ \postsep{}\\ {}
\end{array}
$} & \phantom{$-$}0.21{\scriptsize$
\begin{array}{l} \postsep{***}\\ \postsep{
\includegraphics{624i05.eps}
}\\
\postsep{
\includegraphics{624i01.eps}
}
\end{array}
$} & \phantom{$-$}0.13{\scriptsize$
\begin{array}{l} \postsep{}\\ \postsep{}\\ {}
\end{array}
$} & \phantom{$-$}0.05{\scriptsize$
\begin{array}{l} \postsep{}\\ \postsep{}\\ {}
\end{array}
$} & \phantom{$-$}0.04{\scriptsize$
\begin{array}{l} \postsep{}\\ \postsep{}\\ {}
\end{array}
$} \\
Respiratory disease & $-1.04${\scriptsize$
\begin{array}{l} \postsep{***}\\ \postsep{
\includegraphics{624i07.eps}
}\\
\postsep{
\includegraphics{624i02.eps}
}
\end{array}
$} & $-0.39${\scriptsize$
\begin{array}{l} \postsep{**}\\ \postsep{
\includegraphics{624i06.eps}
}\\
\postsep{
\includegraphics{624i01.eps}
}
\end{array}
$} & $-0.37${\scriptsize$
\begin{array}{l} \postsep{}\\ \postsep{}\\ {}
\end{array}
$} & $-0.69${\scriptsize$
\begin{array}{l} \postsep{**}\\ \postsep{
\includegraphics{624i06.eps}
}\\
\postsep{
\includegraphics{624i01.eps}
}
\end{array}
$} & $-0.71${\scriptsize$
\begin{array}{l} \postsep{***}\\ \postsep{
\includegraphics{624i07.eps}
}\\
\postsep{
\includegraphics{624i02.eps}
}
\end{array}
$} \\
Other infectious disease & $-0.35${\scriptsize$
\begin{array}{l} \postsep{}\\ \postsep{}\\ {}
\end{array}
$} & \phantom{$-$}0.37{\scriptsize$
\begin{array}{l} \postsep{}\\ \postsep{}\\ {}
\end{array}
$} & $-1.14${\scriptsize$
\begin{array}{l} \postsep{*}\\ \postsep{
\includegraphics{624i05.eps}
}\\
\postsep{
\includegraphics{624i01.eps}
}
\end{array}
$} & $-1.72${\scriptsize$
\begin{array}{l} \postsep{***}\\ \postsep{
\includegraphics{624i06.eps}
}\\
\postsep{
\includegraphics{624i02.eps}
}
\end{array}
$} & $-0.89${\scriptsize$
\begin{array}{l} \postsep{*}\\ \postsep{
\includegraphics{624i08.eps}
}\\ {}
\end{array}
$} \\
Traffic injury & $-3.76${\scriptsize$
\begin{array}{l} \postsep{***}\\ \postsep{
\includegraphics{624i07.eps}
}\\
\postsep{
\includegraphics{624i03.eps}
}
\end{array}
$} & $-3.48${\scriptsize$
\begin{array}{l} \postsep{***}\\ \postsep{
\includegraphics{624i07.eps}
}\\
\postsep{
\includegraphics{624i03.eps}
}
\end{array}
$} & $-1.48${\scriptsize$
\begin{array}{l} \postsep{***}\\ \postsep{
\includegraphics{624i07.eps}
}\\
\postsep{
\includegraphics{624i03.eps}
}
\end{array}
$} & $-1.44${\scriptsize$
\begin{array}{l} \postsep{***}\\ \postsep{
\includegraphics{624i07.eps}
}\\
\postsep{
\includegraphics{624i03.eps}
}
\end{array}
$} & $-2.11${\scriptsize$
\begin{array}{l} \postsep{***}\\ \postsep{
\includegraphics{624i07.eps}
}\\
\postsep{
\includegraphics{624i03.eps}
}
\end{array}
$} \\
Suicide & \phantom{$-$}0.25{\scriptsize$
\begin{array}{l} \postsep{}\\ \postsep{}\\ {}
\end{array}
$} & \phantom{$-$}1.06{\scriptsize$
\begin{array}{l} \postsep{***}\\ \postsep{
\includegraphics{624i05.eps}
}\\
\postsep{
\includegraphics{624i01.eps}
}
\end{array}
$} & \phantom{$-$}0.94{\scriptsize$
\begin{array}{l} \postsep{*}\\ \postsep{
\includegraphics{624i08.eps}
}\\ \postsep
{
\includegraphics{624i01.eps}
}
\end{array}
$} & \phantom{$-$}0.80{\scriptsize$
\begin{array}{l} \postsep{*}\\ \postsep{\hspace*{-0.5pt}
\includegraphics{624i08.eps}
}\\
\postsep{
\includegraphics{624i04.eps}
}
\end{array}
$} & \phantom{$-$}0.77{\scriptsize$
\begin{array}{l} \postsep{**}\\ \postsep{\hspace*{-0.5pt}
\includegraphics{624i08.eps}
}\\
\postsep{
\includegraphics{624i04.eps}
}
\end{array}
$} \\
Homicide & $-1.71${\scriptsize$
\begin{array}{l} \postsep{***}\\ \postsep{
\includegraphics{624i05.eps}
}\\
\postsep{
\includegraphics{624i01.eps}
}
\end{array}
$} & $-1.20${\scriptsize$
\begin{array}{l} \postsep{*}\\
\postsep{\hspace*{-0.5pt}
\includegraphics{624i08.eps}
}\\\postsep{\includegraphics{624i04}}
\end{array}
$} & $-1.02${\scriptsize$
\begin{array}{l} \postsep{}\\ \postsep{}\\ {}
\end{array}
$} & $-0.74${\scriptsize$
\begin{array}{l} \postsep{}\\ \postsep{}\\ {}
\end{array}
$} & $-0.4$1{\scriptsize$
\begin{array}{l} \postsep{}\\ \postsep{}\\ {}
\end{array}
$} \\
\hline
\end{tabular*}
\tabnotetext[]{}{Columns represent models, as described in
equation~(\ref{eq:model}) and
Table~\ref{tab:models}.
Rows represent mortality categories.
Table entries are estimates of $100\betaU$, using OLS with states
weighted by the square root of the state population.
Statistical significance is shown using standard OLS errors (black
symbols, top row), error estimates clustered by state (gray symbols,
middle row) and error estimates of \citet{cameron11}, Section~2.2,
with two-way clustering by state and year (gray symbols, bottom row;
red in electronic version).
$^{\ast\ast\ast}P<0.001$, $^{**} P<0.01$, $^{*}P<0.05$,
$^{\tiny +} P<0.1$.}
\end{table}

Table~\ref{tab:methods} shows that the results for age-specific
mortality are also sensitive to model specification.
Model~B1 demonstrates considerable consistency with the 1972--1991
results of \citet{ruhm00}, Table~III, indicating strong procyclical
mortality in all age groups and especially in young adults and middle
age adults.
Our model~L1, which corresponds to a supplementary model for \citet
{ruhm00} and the primary model structure for \citet{miller09}, is in
close agreement with the 1978--2004 results of \citet{miller09}.
In particular, L1 suggests that procyclical mortality is weak or
nonexistent in young adults and middle age adults, and is therefore in
conflict with the conclusions suggested by~B1.
Model~D1 suggests effect estimates which are relatively small, while
being broadly indicative of procyclical mortality across age groups.
Model~HP1$_{100}$ suggests consistent procyclical mortality across age
groups, with smaller effect sizes than~B1.
\citet{ionides12-aoas-sup} show that a Hodrick--Prescott smoothing
parameter of $\lambda=100$ has superior prewhitening properties to
$\lambda=6.25$, and the corresponding regression therefore has more
statistical power to identify cyclical effects.\looseness=-1

From a methodological perspective, the cause-specific mortality results
in Table~\ref{tab:methods} tell a similar story to the age category results.
Traffic fatalities, typically the most clearly procyclical mortality
cause, are highly statistically significant in all analyses.
Procyclical cardiovascular mortality is identified by all models, but
is insignificant for~D1 and~HP1$_{6.25}$.
Model~D1 typically estimates small effect sizes, relative to the other
models in Table~\ref{tab:methods} and relative to previous reports in
the literature: we propose an explanation for this later.
For cancer, models~B1 and~L1 detect a cyclical effect, with opposite signs!
Model~B1 also fails to find a cyclical pattern for suicide, which has
been considered countercyclical in the US [\citet
{luo11,eyer77,ruhm00,tapia05-ije}].
When methodologies disagree on detection of accepted relationships,
they do not provide a firm foundation for investigating new phenomena.
For example, the cyclical behavior of respiratory disease mortality has
previously received relatively little attention, perhaps because it is
somewhat unexpected.
Table~\ref{tab:methods} agrees with other studies [such as \citet
{miller09}] in detecting a clear procyclical pattern of mortality due
to respiratory disease.

The state clustered standard errors in Table~\ref{tab:methods}
generally produce similar conclusions to the OLS standard errors, with
some important exceptions.
For models~D1, HP1$_{6.25}$ and HP1$_{100}$, state clustered standard
errors are generally similar in magnitude to OLS standard errors
(results not shown).
This is to be expected when residual autocorrelation is small, and in
this case state clustered standard errors may be less reliable than the
usual OLS standard errors [\citet{kauermann01}].
For models~B1 and~L1, many large effect sizes remain significant
despite substantially inflated clustered standard errors.
Conclusions about the effects on suicide and cardiovascular disease are
noticeably sensitive to the use of state clustered standard errors.
These two mortality categories are also identified in
Section~\ref{sec:diagnostics} as having inconsistent effects between states.
Inconsistency between states leads to relatively large state clustered
standard errors, since these error estimates assess uncertainty by
quantifying variability between states rather than between time points.

The five models in Table~\ref{tab:methods} emphasize relationships at
different ranges of frequencies.
The synchronous fluctuations of many macroeconomic variables around
their trends, known as business cycles, are of irregular duration and
have a power spectrum spread broadly over a wide range of frequencies
[\citet{canova98}].
It need not be the case that all frequencies of macroeconomic
fluctuations are equally associated with population health.
The range of frequencies at which the statistical evidence for cyclical
associations is greatest could, potentially, differ from the range at
which the public health consequences are greatest.
One way to study these issues is through spectral analysis [\citet
{tapia08-jhe}], but here we simply interpret the frequency-domain
behavior of the specified regression models [\citet{ionides12-aoas-sup},
Section S3].
Model~B1 performs the least detrending and therefore places the most
emphasis on low frequency associations.
This leads to some large effect estimates, matched with increased
uncertainty (which can be viewed as larger standard errors or unknown biases).
Model~HP1$_{6.25}$ emphasizes a range of frequencies intermediate
between D1 and HP1$_{100}$, and the results for HP1$_{6.25}$ are
generally intermediate between these two analyses.
Model~D1 emphasizes the highest frequencies, to such an extent that
some cyclical relationship becomes obscured.
Macroeconomic fluctuations involve complex relationships between many
variables [\citet{canova98}] and so it may be unreasonable to expect any
single economic measure to capture reliably, at high frequencies, the
relationship to health outcomes.
Traffic injuries might be expected to have a relatively clean
high-frequency relationship to economic activity (proxied by
unemployment in our models), as there is an obvious and immediate
causal mechanism.
However, even for traffic mortality, the parameter estimates for
models~D1 and~HP1$_{6.25}$ are smaller than for the other models.

Inasmuch as equation~(\ref{eq:model}) is valid, all the estimation
methods result in unbiased effect estimates: the weighting of
frequencies in the estimation procedure affects the variability of the
OLS estimate but not its bias.
However, in practice, one cannot expect any model to be equally
appropriate over all time scales.
Investigating the time scales at which the model applies is therefore
an integral part of data analysis.
Model~HP1$_{100}$ emphasizes an intermediate range of frequencies and
is seen to provide the clearest statistical evidence for cyclical mortality.

If cyclical mortality were to exist only in the context of fluctuations
around a trend, then it would have no long term consequences, since
above-trend and below-trend fluctuations necessarily cancel out in the
long run.
Alternatively, if cyclical mortality were present in macroeconomic
fluctuations occurring over a decade or longer, one should consider the
cyclical effects at least partly responsible for observed health trends
on these time
scales.
The indications from model~B1 that procyclical mortality may be even
stronger at low frequencies support this second interpretation.

%
\begin{figure}

\includegraphics{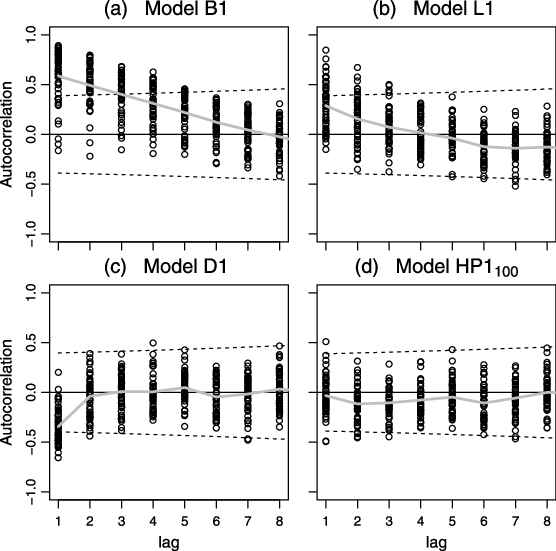}

\caption{Autocorrelation of the residual in four models for total mortality.
Points show the sample autocorrelation for each state at each lag.
The dashed lines are at ${}\pm t_{n-2}\{n-2+t_{n-2}^2\}^{-1/2}$, where
$t_{n-2}$ is the $97.5$
percentile of the $t$ distribution on $n-2$ degrees of freedom and $n$
is the number of pairs
of time points available to compute the sample autocorrelation at each lag.
If the residual series were temporally uncorrelated, approximately 95\%
of the points should
lie between the dashed lines [\citet{moore99}, Section~10.2].
The gray solid line graphs the mean sample autocorrelation at each lag.}\label{fig:autocor}
\end{figure}

\section{Diagnostic analysis}\label{sec:diagnostics}
The spatiotemporal dependence of the regression residuals affects the
appropriate choice of model specification, the suitability of parameter
estimation methodologies and the evaluation of uncertainty in the
resulting estimates.
Figure~\ref{fig:autocor} shows the temporal autocorrelation of the
residuals for each state at each lag.
We see that there is strong positive autocorrelation for model~B1, at
short lags.
This positive autocorrelation is reduced, but still substantial, for
model L1.
The autocorrelation for model D1 becomes significantly negative at lag
1, as might be expected from applying a differencing operation.
There is some indication of mild negative autocorrelation after lag 1
for model HP1$_{100}$, but this model shows relatively minor deviation
from the expected behavior of uncorrelated residuals.\vadjust{\goodbreak}

%
\begin{figure}

\includegraphics{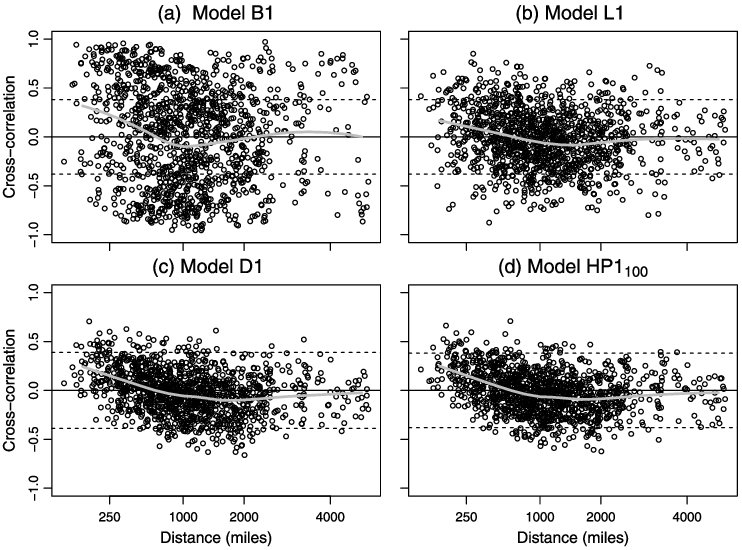}

\caption{The cross-correlation between residuals for each pair of
states, plotted against distance between population-weighted state
centers (from the 2000 census) in four models for total mortality.
The dashed lines are at ${}\pm t_{n-2}\{n-2+t_{n-2}^2\}^{-1/2}$, where
$t_{n-2}$ is the $97.5$ percentile of the $t$ distribution on $n-2$
degrees of freedom, and $n=27$ (for \textup{B1}, \textup{L1}, \textup{HP1}$_{100}$) or $n=26$ (for \textup{D1}).
If the residual series were spatiotemporally uncorrelated,
approximately 95\% of the points should lie between the dashed lines
[\citet{moore99}, Section~10.2].
The actual percentages for models~\textup{B1}, \textup{L1}, \textup{D1} and \textup{HP1}$_{100}$ are 46.1\%, 79.3\%, 90.9\% and 91.3\%, respectively.
The gray solid line shows a local linear regression fit to these
cross-correlations, implemented using the \texttt{loess} function in
\textup{R2.15.0}, with the default parameter settings.}\label{fig:cor-dist}
\end{figure}

Similar patterns emerge when investigating spatial correlation.
Figure~\ref{fig:cor-dist} shows the sample correlations between the
time series of residuals for all 1225 $(= 50\times49/2)$ pairs of states.
Models B1 and L1 show considerably more variability in the sample
correlation that is consistent with spatiotemporally uncorrelated residuals.
The sample autocorrelations of the residuals are necessarily centered
near zero, due to the inclusion of year effects.
The lack of a substantial spatial pattern suggests that dependence
between neighboring states is not a major concern.
The increased spread is another indication of temporal correlation:
independent sequences which each have positive marginal temporal
autocorrelation typically have sample cross-correlation with mean zero
but greater variability than temporally uncorrelated sequences.
Models D1 and HP1$_{100}$ have a spread of sample cross-correlations
which is approximately consistent with spatiotemporally uncorrelated residuals.
The lower variability for models D1 and HP1$_{100}$ reveals a small
pattern of positive correlations between residuals of states in close proximity.
It would be surprising if no such phenomenon existed, but we see here
that the effect is rather weak.
Most of the cross-correlation of fluctuations in mortality between
states is removed by the inclusion of the national year effect $\betaY_t$.
If these year effects are not included (i.e., in models of subtype 2, 3 or~4),
a plot analogous to Figure~\ref{fig:cor-dist} shows consistently
positive cross-correlations across all geographic distances (results
not shown).

%
\begin{figure}

\includegraphics{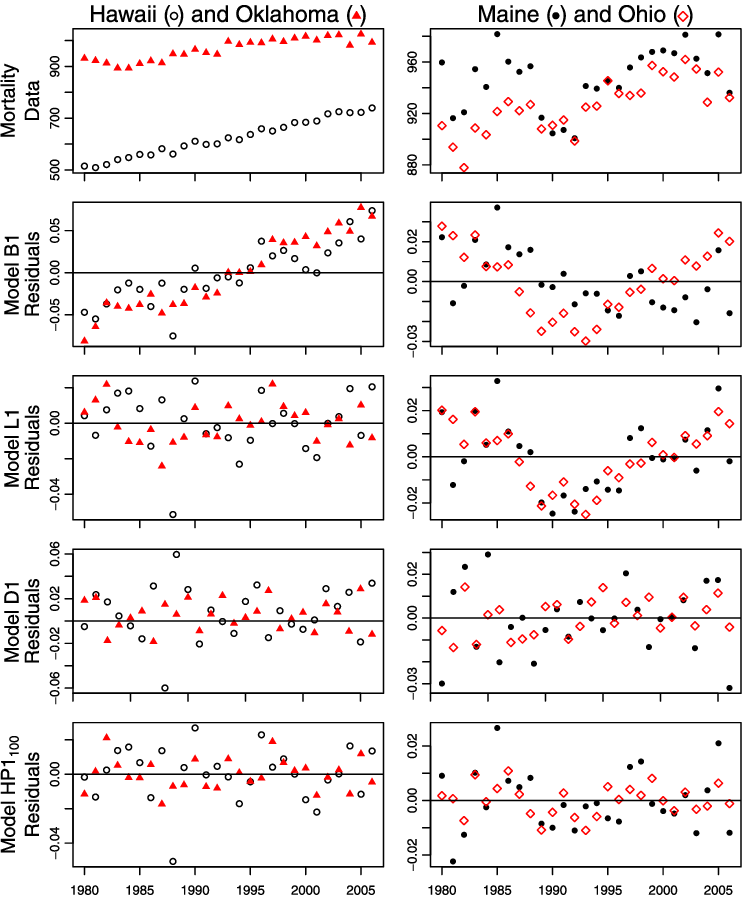}

\caption{Residual time plots for four states. The top row graphs total
state mortality,
and subsequent rows graph residuals for each of four models.}\label
{fig:state-example}
\end{figure}

Residuals can also be investigated by examination of the time plots for
each state.
Some representative time plots are shown in Figure~\ref{fig:state-example}.
This figure reinforces the conclusion that OLS estimation of the basic
model is a questionable practice, since the underpinning model
assumptions are violated for almost all states.
The linear trend model is sometimes adequate (e.g., Hawaii and
Oklahoma) and sometimes not (e.g., Maine and Ohio).
Both differencing and HP detrending remove systematic trends from the
time series of residuals.

\begin{figure}[t!]

\includegraphics{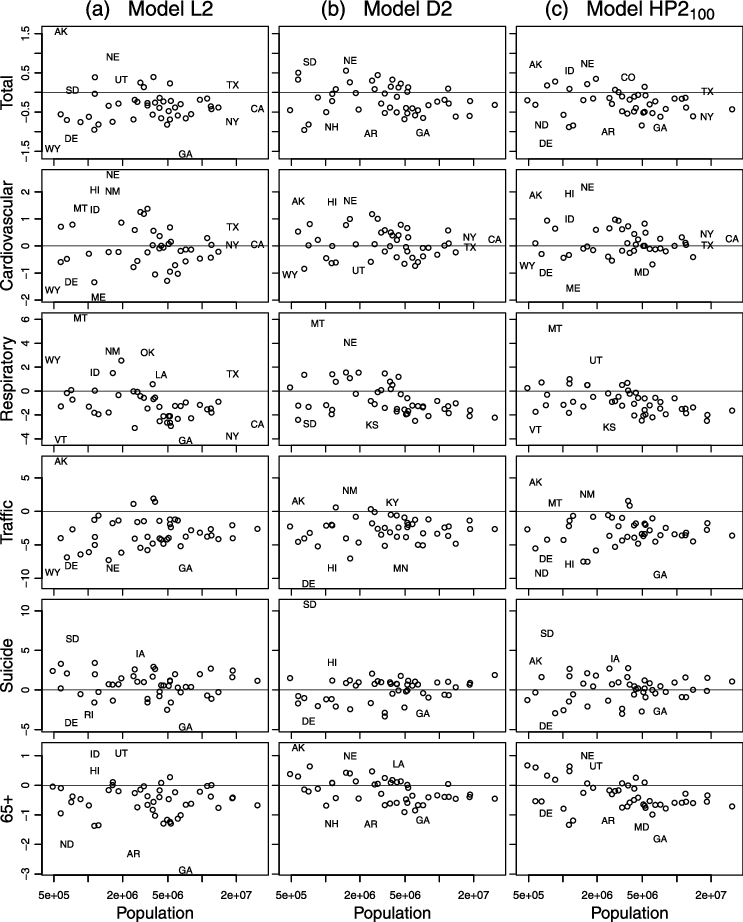}

\caption{State-specific effects of unemployment on mortality.
Columns correspond to models, as specified in Table~\protect\ref
{tab:models} and equation~(\protect\ref{eq:model}).
Rows correspond to mortality categories.
The estimate of $100\betaU$ from fitting the model to a single state is
plotted against the population of the state.
Each state is represented either by its two letter abbreviation or by
an open circle.}\label{fig:state-effects}\vspace*{6pt}
\end{figure}

The conclusion from these diagnostic investigations is that, among
these alternatives, model HP1$_{100}$ unambiguously comes closest to
satisfying the model assumptions for a standard linear model analysis.
As another criterion to compare model specifications, we compared the
consistency of the estimated cyclical mortality effects between states.
A robust relationship between macroeconomic fluctuations and mortality
might be expected to demonstrate consistent results in separate
state-by-state time series analyses.
We explored the stability of the panel model effect estimates across
states by estimating the unemployment effect on mortality using data
for one state only, that is, the model in equation (\ref{eq:model}) for
a single fixed value of the state label $i$.
For a state-by-state analysis, one cannot estimate fixed year effects,
but one can still estimate models of subtypes~2--4.
The results for subtype~2 are plotted in Figure~\ref
{fig:state-effects}, from which we observe that HP2$_{100}$ provides
the greatest consistency between states, closely followed by D2.
For example, the standard error of the 50 state-specific estimates of
$100\betaU$ for total mortality is 0.53 for~L2, 0.45 for~D2, and 0.43
for~HP2$_{100}$.
National fluctuations in mortality unrelated to the economy, perhaps
due to infectious disease or climate, are not controlled for in model
subtype 2.
Some mortality categories nevertheless demonstrate consistent
state-by-state effects, especially for the larger states.
As might be expected, there is typically greater variation in the
estimated effects for smaller states.
From Figure~\ref{fig:state-effects}, we see that the effects for total
mortality, respiratory disease, traffic injuries and ages 65$+$ have
consistent signs in all (or almost all) of the larger states.
Perhaps surprisingly, suicide and cardiovascular disease show only weak
patterns in the state-by-state analysis despite the evidence for
overall cyclical behavior from the full panel analysis (Table~\ref
{tab:methods}, column HP1$_{100}$).

%


\section{Conclusions} \label{sec:discussion}

We have seen that the choice of model can have considerable influence
on panel analysis of the associations between fluctuations in mortality
and macroeconomic variables.
These influences are simultaneously a concern, a~challenge and an opportunity.
The concern is that, unless a methodological consensus is found,
scientific claims which are sensitive to choice of methodology must
remain unresolved.
The challenge is to establish statistical procedures which objectively
assess the strengths and weaknesses of different analyses, and so
disambiguate the conclusions.
Overcoming this challenge will give an opportunity to improve
understanding of the phenomenon of procyclical mortality.
A historical precedent for methodological introspection in this
research area is the examination and eventual rejection of the methods
employed by Harvey Brenner.
Indeed, panel analyses have played an important role in clarifying the
evidence for overall procyclical mortality.
However, we have shown that previous panel approaches have limited
capability to identify more subtle components of the cyclical
effect.\looseness=1

It is well known that positive temporal autocorrelation [\citet
{bertrand04}] and positive spatial cross-correlation
[\citet{layne07}]
typically cause OLS standard errors for panel models to be
anti-conservative (i.e., inappropriately small).
Underestimated standard errors lead to overestimated statistical
significance, and hence the detection of spurious relationships.
Clustering standard errors by state helps to resolve this issue, but
these robust standard errors fail to correct for dependence between states.
Clustering standard errors by state and year gains additional
robustness, with the cost being increased variability in the standard
error estimates.
In addition, the OLS regression coefficient estimates remain
inefficient (if unmodeled trends are considered random variables) or
biased (if unmodeled trends are considered as fixed effects).
We have shown that nonlinear detrending can be employed to fix these
methodological shortcomings in the context of investigating cyclical mortality.

The study of cyclical mortality fluctuations is sensitive to these
methodological issues because relatively small effects, which are hard
to unravel from other background sources of variability, can
nevertheless have substantial consequences at the population level.
The larger and clearer the effect, the less sensitive its detection
should be to the details of the statistical methodology used to
investigate it.
However, understanding the overall pattern requires investigating which
subpopulations and mortality causes are involved.
Inevitably, one seeks to press to the limits of the available data and
statistical methodology.

We have proposed a resolution to the differing accounts of
age-dependency for procyclical mortality.
Our preferred specification (Table~\ref{tab:methods}, column
HP1$_{100}$) suggests that the effect is relatively uniform across
ages, which has attractive conceptual simplicity.
There may be no reason a priori to expect age uniformity.
In particular, individuals in the 65$+$ age category are predominantly
out of the workforce: they are therefore largely unaffected by some
potential mechanisms such as extra hours of work, or fewer hours of
sleep, during economic expansions.
The 20--44 age category has the largest estimated effect for model
HP1$_{100}$, yet, according to the spatiotemporal clustered errors,
this age group is the only one in which the association is
statistically insignificant.
Other lines of reasoning, including the spatiotemporal clustered errors
for other choices of the Hodrick--Prescott smoothing parameter, and
other choices of standard error for model HP1$_{100}$, suggest adequate
statistical evidence for this association.

Our results for cause-specific mortality (Table~\ref{tab:methods},
column HP1$_{100}$) give a single set of figures consistent with
previous analyses but without the occasional peculiarities that are a
hallmark of misspecified models.
For example, the models B1 and L1 suggest macroeconomic associations
for cancer with differing signs.
The statistical significance of cancer for model~B1 disappears when
using clustered standard errors; for L1, the countercyclical
association remains significant.
\citet{miller09} found a countercyclical association with cancer (with
unspecified statistical significance) consistent with the similarity of
their model specification to~L1.
\citet{tapia05-ije} found a procyclical association in the US for
1945--1970, but not in other time intervals.
The long lag times involved in the chronic development of cancer are
hard to reconcile with an unlagged cyclical relationship.
However, it is entirely possible that external factors could be
associated with acute complications resulting in death of cancer patients.
This possibility is self-evident for cardiovascular disease, where
acute cardiovascular failures are associated both with chronic disease
development and external stress.

Cardiovascular disease and cancer are the two foremost causes of death
in developed countries, and the cyclical behavior of cardiovascular
mortality has therefore attracted considerable attention [\citet{ruhm07}].
Cardiovascular disease mortality has a relatively small procyclical
signature over the 23 developed countries in the Organization for
Economic Cooperation and Development (OECD) studied by \citet{gerdtham06}.
In some countries, such as Japan [\citet{tapia08-japan}], the
procyclical signature of cardiovascular disease mortality seems to be
strong; in others, such as Germany [\citet{neumayer04}], it seems to be
negligible.
In Sweden there is some evidence for a countercyclical effect [\citet
{svensson08,tapia11}].
In the US, Table~\ref{tab:methods} reconfirms the conclusions of
\citet
{ruhm00} and \citet{miller09} that the dominant behavior of
cardiovascular disease is procyclical.
However, we found in Figure~\ref{fig:state-effects} that this result is
not strongly consistent at the level of individual states.

The unambiguous evidence for procyclical respiratory mortality requires
further investigation.
This phenomenon has been noted in other studies of developed countries
[\citet{eyer77,ruhm00,neumayer04}, Tapia Granados (\citeyear{tapia05}), \citet{gerdtham06,miller09}], but it
has become further clarified by the statistical methods we have employed.
Specifically, we have shown the strong consistency between individual
states, and we have employed methods that minimize the risk of
identifying spurious relationships.
Our data cannot readily reveal how mechanisms such as air quality
(pollution) and weakened immune status (increased infectious disease
transmission) may combine to produce this procyclical effect.
Respiratory disease, as categorized in ICD9/10, is not necessarily
infectious but does include influenza and pneumonia, which are
responsible for substantial mortality in old age.
Infectious diseases provide a potential avenue by which those outside
the workforce suffer procylical mortality, since collective resistance
plays a substantial role in controlling disease spread [an effect known
as \textit{herd immunity} in epidemiology; \citet{bonita06}].
Overwork and a reduction in healthy behaviors during economic booms
could lead to a population with weaker overall health and therefore
greater transmission of pathogens.
Increased travel, associated with increased economic activity, provides
another potential mechanism for increased transmission of pathogens.

Previous studies [\citet{ruhm00,miller09}] have found that homicides
oscillate procyclically.
This result may appear counterintuitive and, to our knowledge, it has
not been fully explored.
According to our specification~HP1$_{100}$ (and also~D1
and~HP1$_{6.25}$) in Table~\ref{tab:methods}, there is no clear
evidence that homicides are correlated with the business cycle.
Inasmuch as the data support procyclical homicide, this is based on the
models~B1 and~L1 which place more emphasis than~HP1$_{100}$ on
longer-term variation.

Our analyses provide weak support for an overall countercyclical nature
of suicide in the US, consistent with the conclusions of \citet{luo11}.
A cyclical effect on suicide might be intuitively unsurprising, but the
direction of the effect is not consistent between countries.
For example, suicide in Japan is strongly countercyclical [\citet
{tapia08-japan}], whereas in Germany and Finland there is evidence for
procyclical suicide
[\citet{neumayer04,hintikka99}].
No dominant pattern was found in a study of OECD data [\citet{gerdtham06}].
Figure~\ref{fig:state-effects} suggests that the cyclical behavior of
suicide is inconsistent between states.
This conclusion is supported by the diminished significance of the
overall countercyclical effect once the standard errors are clustered
by state.\looseness=1

Debate about individual components of cyclical mortality, and remaining
uncertainty about specific causal mechanisms, should not obscure the
main achievement of recent research in this area.
There is now overwhelming evidence that downturns in economic activity
have not had overall adverse health consequences at the population
level, in the recent past of developed countries with market economies.
Groups of individuals adversely affected by phenomena associated with
economic booms and busts deserve assistance.
At earlier stages of socioeconomic development, economic growth may
have substantial health benefits above and beyond other factors such as
public health programs and education [\citet{pritchett96}].
However, the government's responsibility to consider the net public
health consequences of its actions [\citet{childress02}] cannot be used
as a moral argument for pro-growth economic policies in the US and
similar countries.
Other moral obligations relevant to macroeconomic policy include the
protection of individual liberties, environmental stewardship and
homeland security.
Future public policies will require trade-off between economic growth
and other objectives, and evidence-based assessment of the positive and
negative consequences of economic growth should inform this
debate.\looseness=1

\section*{Acknowledgments}
Helpful
suggestions were provided by the Editor, an Associate Editor and an
anonymous referee.

\begin{supplement}[id=suppA]
\stitle{Supplement to ``Macroeconomic effects on mortality revealed by
panel analysis with nonlinear trends''}
\slink[doi]{10.1214/12-AOAS624SUPP} 
\sdatatype{.pdf}
\sfilename{aoas624\_supp.pdf}
\sdescription{We present supplementary material on:
(i) interpretation of detrending choices;
(ii)~data analysis
for additional detrending choices;
(iii) prewhitening as a goal for selecting the Hodrick--Prescott
smoothing parameter.}
\end{supplement}

%
%


\printaddresses

\end{document}